\documentclass[12pt]{article}         
\usepackage{amsfonts}
\usepackage{amsmath}
\usepackage{latexsym}
\usepackage{color}
\usepackage{cite}\normalsize
\usepackage{graphicx}[12pt]
\newcommand{\bm}[1]{\mbox{\boldmath$#1$}}

\newcommand{\bea}{\begin{eqnarray}}
\newcommand{\eea}{\end{eqnarray}}

\def\be{\begin{equation}}
\def\ee{\end{equation}}
\def\bea{\begin{eqnarray}}
\def\eea{\end{eqnarray}}

\title{On the use of the autonomous Birkhoff equations  in  Lie series perturbation theory}
\author{T.S. Boronenko\thanks{boron@tspu.edu.ru} \\
Tomsk State Pedagogical University, 634061 Tomsk, Russia}
\date{ }
\begin{document}
\maketitle
\begin{abstract}
In this article,  we present the Lie  transformation algorithm  for  autonomous Birkhoff systems. 
Here, we are referring to Hamiltonian systems that obey a symplectic structure
of the general form.
The Birkhoff equations are  derived from the  linear first-oder Pfaff-Birkhoff
variational 	principle, which is more general than the Hamilton principle.
The use of 1-form in the formulation of the equations of motion in dynamics
makes  the Birkhoff method   more universal and flexible. Birkhoff's equations  have a tensorial character, so their 
form is independent of the coordinate system  used.
Two examples of normalization in the restricted three-body problem  are given to illustrate the application of the 
algorithm in perturbation theory.
The efficiency of this algorithm for  problems of asymptotic integration in dynamics is discussed for the case
where there is a need to use non-canonical variables in the phase space.
\vspace{2pt}

\end{abstract}

\section{Introduction}\label{sec1}
Birkhoff's autonomous equations   are the Hamilton equations expressed in terms of non-canonical variables in phase space.
We assume that a local coordinate transformation exists and is not  explicitly dependent  on time, and  equations  can always be
reduced to canonical form ( Darboux theorem; see e.g., \cite{Arnold}). In this article, we show that  the Birkhoff
equations can be useful in problems of Celestial Mechanics, and in particularly in perturbation theory.

The construction of analytical models   in high-order  perturbation theory  requires rather cumbersome calculations, which 
 is practically impossible without modern computer algebra systems. The  Lie transformation algorithm (\cite{Hori,Deprit})
 is  best suited for algebraic  manipulations of this kind.
 Despite the seeming simplicity of the  method, in the process of solving practical problems,  it is usually required 
to perform  a large number  of transformations of different systems of variables. For example, theoretical constructions
in problems of perturbation theory are more easily  performed in canonical variables, but in practice it is not always 
convenient to use them.

There are more serious problems as well. Hamiltonian and Lagrangian   formulations of  dynamics are associated with some constraints.
In the former,  the variables must be canonical, in the latter,  transformations of variables can be done only  in the configuration space.

 In this context, a method which is based on Pfaffian or linear differential form is more flexible (\cite{Blimovitch, Broucke}). 
Pfaff's  equations or
the 'associated Pfaff equations'   were  published by \cite{Pfaff}.  However, in this publication,  these equations were not considered    in the quality of the   analytical dynamic equations.
These properties of equations were identified
 in full by \cite {Birkhoff}.  Birkhoff  also showed that the equations could be derived from a variational principle.
 For these reasons,  Santilli \cite{Santilli} proposed  calling the equations 'Birkhoff equations'.
 Variational principle came to be called the Pfaff-Birkhoff integral variational principle.  
 This terminology has naw been  adopted by a number of authors (e.g., \cite{Sun, Zhang}). 
 In our work, we use the study  by  \cite{Santilli},
which showed that the Birkhoff equations preserved the Lie algebra character of the canonical Hamilton equations, and that these
equations have the  most general symplectic structure in  local coordinates.

The application of the Pfaffian or 1-form to the formulation of equations of motion in
Celestial  Mechanics was investigated by \cite{Broucke},  who showed that the method enabled the use of a very broad class
of variables in constructing  equations of motion - in particular, the classical Kepler elements.  In this case, the 
equations of motion are the Lagrange planetary equations. The possibility of using  Kepler's elements as the orbital 
coordinates was considered by \cite{Abraham}.

The concept, in accordance
with which all operations of the Lie transformation algorithm for certain problems of Celestial Mechanics can be
made in the Kepler elements, was first outlined
by \cite{Kholshevnikov}. The Lie transformation algorithm, presented in the Kepler elements, 
was used to construct the analytical theory of Phoebe, the ninth satellite of Saturn \cite{Boronenko_Shmidt}.

This paper provides a general method for constructing the Lie transformation algorithm  for the autonomous Birkhoff 
systems, and here we  consider only the formal aspects of dynamics that
have no relation to the issues of convergence or divergence of series and to  the problems of equilibrium.

The paper is organized as follows.

Section \ref{sec2} provides a description of the method. Here, we  briefly describe  the  Lie transformation
algorithm for Hamiltonian systems for the case where the solution  of the problem requires a great number of approximations.
We then  introduce special variables in  phase space and show the possibility of using the autonomous Birkhoff equations  
in the Lie series  perturbation theory.

 In Sect.\ref{sec3}  we present two examples  using the algorithm in problems of perturbation theory.

In the first  example, we consider the satellite case of the spatial restricted three-body problem.
The averaging method based on Lie  transformations of Birkhoffian system is  proposed for the case where
the expansions for short-period perturbations are presented  in powers of $ m $ (ratio of mean motions of
Sun  and satellite), but in closed form with respect to eccentricity and inclination.

In the second example, we present an  analytical solution of the restricted three-body problem using the Delaunay arguments.
Unlike the previous example, here we deal with an explicit expression of the perturbing function in terms of
the  mean anomaly of the satellite. Therefore, all the  expansions considered now   include power series in the eccentricity
of the  satellite orbit.
We applied  a similar algorithm in an  earlier work for the construction of an analytical theory of motion of Phoebe,
the ninth satellite of Saturn  \cite {Boronenko_Shmidt}.  In this paper we derived this algorithm in the context of the
Birkhoff theory. 

Section \ref{sec4} contains discussion of the results and our conclusions.

\section{Description of the method}\label{sec2}
\subsection{Lie transformation  algorithm }
\label{sec 2.1}
Here we provide  a brief description of the Lie transformation  method applied to  canonical perturbation theory in the case
of high-order computations. In this algorithm, all operations are based on the Lie series \cite{Hori, Ferraz-Mello, Kholshevnikov_K}.

 Let us consider the Hamiltonian in the form of  formal truncated series in $\tau$:
\begin{equation}
\label{am1}
H(\eta)= H_{00}(\eta)+\tau H_{01}(\eta)+\dots+\tau^m H_{0m},
\end{equation}
where through $\eta$, two sets are denoted:  $\eta_1,\dots\,\eta_l$ are generalized coordinates and $\eta_{l+1},\dots\,\eta_n$
are generalized momenta. Thus, the dimension of the phase space is given by $n=2l$. We assume that $H(\eta)$ 
is a smooth time-independent function,  $\tau$
is a constant small parameter and $H_{00}$ is the Hamiltonian of an integrable system. The corresponding canonical equations
have the form:
\begin{equation} \label{am2}
\dot{\eta}_i=\frac{\partial{H(\eta,\tau)}}{\partial \eta_{i+l}}, \ \dot{\eta}_{i+l}=-\frac{\partial{H(\eta,\tau)}}{\partial \eta_{i}},\  i=1,\ldots,l.
\end{equation}
We introduce a transformation $\eta\rightarrow\tilde{\eta}$ that is determined by the Lie series.
The concept of Lie series follows from the solution of the Cauchy problem, which can be formulated  in the following  way
(see e.g., \cite{Kholshevnikov_K}):
\begin{eqnarray} \label{am3}
\dot{\eta}_i=\frac{\partial W}{\partial \eta_{i+l}},\ \dot{\eta}_{i+l}=-\frac{\partial W}{\partial \eta_{i}},\nonumber\\
\eta_i|_{t=0}=\tilde{\eta}_i,\;\:\eta_{i+l}|_{t=0}=\tilde{\eta}_{i+l},\nonumber\\
i=1,\ldots,l,
\end{eqnarray}
where $W=W(\eta,\tau)$ is some analytic function in a neighborhood of the initial point $\tilde{\eta}_0=
(\tilde{\eta}_{01},\ldots,\tilde{\eta}_{0n})$.
The solution of (\ref{am3})  at $t=\tau$ is determined by the Lie series
\begin{equation} \label{am4}
\eta_j=\sum^{\infty}_{m=0}\frac{\tau^m}{m!}L^m_W \tilde{\eta}_j=exp(\tau L_W)\tilde{\eta}_j,\ j=1,\ldots,n.
\end{equation}
The general solutions  (\ref{am4}) is a one-parameter group of canonical transformations.  In accordance with
 the terminology  adopted in the theory of  Lie groups, the function $W$ is called the Lie generating function or 
the Lie  generator of the group (see e.g., \cite{Ferraz-Mello}). In (\ref{am4}) the Lie generator represents the Poisson bracket
 of the form:
\begin{equation} \label{am5}
L_W=\sum^l_{i=1}\left(\frac{\partial W}{\partial {\tilde{\eta}}_{i+l}} \frac{\partial}{\partial {\tilde{\eta}}_{i}}-
\frac{\partial W}{\partial {\tilde{\eta}}_{i}} \frac{\partial}{\partial {\tilde{\eta}}_{i+l}}\right).
\end{equation}

Let us assume that the function W  is represented as a truncated series:
\begin{equation}
\label{am6}
W({\eta})= \tau W_{01}(\eta)+\dots+\tau^m W_{0m}.
\end{equation}
The transformation of the original Hamiltonian $H(\eta)$  can then be performed using the following recursive algorithm:
\begin{eqnarray}
\label{am7}
\widetilde{H}_{0n}&=&\sum_{i=0}^n H_{i,n-i},\ n=0,\dots,m;\nonumber\\
H_{i,n-i}& =&\frac{1}{i}\sum_{\rho=0}^{n-i} \left\{H_{i-1,n-i-\rho}\;,\; W_{0,\rho+1}\right\}, \ i\neq{0},
\end{eqnarray}
where the expression in curly brackets is a Poisson bracket.
We assume that the Hamiltonian is not dependent on time, and this allows us to write  $\widetilde{H}(\widetilde{\eta})=H(\eta)$.
A new Hamiltonian is defined as follows:
\begin{equation}
\label{am8}
\widetilde{H}= \widetilde{H}_{00}+\tau \widetilde{H}_{01} +\dots+\tau^m \widetilde{H}_{0m}.
\end{equation}

The recursive algorithm (\ref{am7})  is easy to demonstrate with the following triangle:
\begin{eqnarray}\label{am9}
& H_{00} & H_{01}  \;\;  H_{02} \;\;  H_{03} \ldots \nonumber \\
& H_{10} & H_{11}  \;\;\  H_{12}  \ldots \nonumber \\
& H_{20} & H_{21}  \ldots \nonumber \\
& H_{30} & \ldots
\end{eqnarray}

In the future, we will use the notation for the variables $\eta$\ without tilde, because in the Lie transformation
algorithm,  new or old variables are generally defined from the context.

It is known (see e.g., \cite{Ferraz-Mello}) that  Lie transformations  have the following general properties. 
These transformations are
defined as  infinitesimal canonical transformations. To obtain the inverse transformation, we must reverse the sign 
of the function $W$.
Transformations of variables and functions of these variables are performed by the same algorithm (\ref{am9}).

The above-described method offers a systematic approach to the problem of the separation of variables in the differential 
equations of the perturbation theory.
 This algorithm is especially effective in combination with averaging methods in Hamiltonian systems.
Such approaches to solving the problems of  perturbation theory are well known in celestial mechanics --
for example, the method of Delaunay,
Von Zeipel, and  Kolmogorov-Arnold.
 These methods  use  the averaging principle and the concept of dividing motion
into slow drift and fast oscillations  (see e.g., \cite{Brumberg}). 

In developing similar algorithms, it is sometimes preferable to use the orbital elements or their functions  
as variables in the phase space. In the next section we  will show that all  operations of the Lie transformations
for Hamiltonian systems can be performed with the help of some special variables, which are non-canonical.

\subsection{ Introduction of the special variables in phase space}
\label{sec2.2}

Let us consider some function $f_{ij}(\eta)\equiv H_{ij}(\eta)$ from triangle (\ref{am9}). The Lie transformation algorithms 
are reduced to the successive calculation of the Poisson brackets of the following form:
\begin{equation} \label{am10}
L_k\:f_{ij}=\sum^l_{c=1}\left(\frac{\partial W_{0k}(\eta)}{\partial {\eta}_{c+l}} \frac{\partial f_{ij}(\eta)}{\partial {\eta}_{c}}-
\frac{\partial W_{0k} (\eta)}{\partial {\eta}_{c}} \frac{\partial f_{ij}(\eta)}{\partial {\eta}_{c+l}}\right),\ i+j=k,
\end{equation}
where $W_{0k}$ are coefficients  in the truncated series  (\ref{am6}). It is assumed that all the functions $W_{0k}(\eta)$  and
$f_{ij}(\eta)$ are sufficiently  smooth.

Now let us suppose that the orbit in  phase space is determined by the values of some suitable  parameters   $\theta_1,\ldots,\theta_n$, which we accept  as new variables.  We then  enter the following transformation:
\begin{eqnarray}
\label{am11}
 \eta_p=\eta_p(\theta_1,\ldots,\theta_n),\ \ p=1,2,\ldots,n.
\end{eqnarray}
It is assumed that   transformations (\ref{am11}) are analytic in their region of definition. In addition, we assume that all
  transformations are regular, i.e., their Jacobian does not vanish:
\begin{eqnarray}
\label{am12}
 J=\frac{\partial{(\eta_1,\ldots,\eta_n)}}{\partial{(\theta_1,\ldots,\theta_n)}}\neq 0.
\end{eqnarray}
As a consequence, the transformations (\ref{am11})  are invertible in their region of definition, with the
inverse transformations given by:
\begin{eqnarray}
\label{am13}
 \theta_p=\theta_p(\eta_1,\ldots,\eta_n).
\end{eqnarray}
 After performing the substitution (\ref{am11}),  the Lie generator   (\ref{am10}) takes the form:
\begin{eqnarray}
\label{am14}
L_{k}f_{ij}(\theta)=\sum_{p=1}^{n}\frac{\partial f_{ij}(\theta)}{\partial \theta_p} \sum_{s=1}^n \frac{\partial W_{0k}(\theta)}{\partial \theta_s} \sum_{c=1}^l\left(\frac{\partial \theta_{p}}{\partial \eta_{c+l}} \frac{\partial \theta_{s}}{\partial \eta_{c}}- \frac{\partial \theta_{p}}{\partial \eta_{c}} \frac{\partial \theta_{s}}{\partial \eta_{c+l}}\right).
\end{eqnarray}
We introduce the Poisson brackets
\begin{equation} \label{am15}
\left\{\theta_p,\theta_s\right\}=\sum_{c=1}^l\ \left(\frac{\partial \theta_{p}}{\partial \eta_{c+l}} \frac{\partial \theta_{s}}{\partial \eta_{c}}- \frac{\partial \theta_{p}}{\partial \eta_{c}} \frac{\partial \theta_{s}}{\partial \eta_{c+l}}\right).
\end{equation}
Now, we can rewrite the expression (\ref{am14})  as follows:
\begin{equation} \label{am16}
L_{k}f_{ij}(\theta)=\sum_{p=1}^{n}\left(\sum_{s=1}^{n} \left\{\theta_{p},\theta_{s} \right\}\frac{\partial W_{0k}(\theta)}{\partial \theta_s}\right) \frac{\partial f_{ij}(\theta)}{\partial \theta_p},
 \end{equation}
or
\begin{equation} \label{am17}
L_{k}f_{ij}(\theta)=\sum_{p,s=1}^{n}\left( a_{ps}(\theta)
\frac{\partial W_{0k}(\theta)}{\partial \theta_s}\right)
 \frac{\partial f_{ij}(\theta)}{\partial \theta_p},
 \end{equation}
where $a_{ps}(\theta)=\left(\left\{\theta_{p},\theta_{s} \right\}\right)$  is a skew-symmetric matrix of Poisson brackets.
The Lie generator can also be represented as follows:
\begin{equation} \label{am18}
L_{k}f_{ij}(\theta)=\sum_{p=1}^n W_{pk}\frac{\partial{f_{ij}}}{\partial \theta_p},\ \ \  (k=i+j).
\end{equation}
Here, $ W_{pk} $ are elements of the matrix $\Psi=\left(W_{pk}\right), (p=1,\ldots,n;k=1,\ldots,m)$,
 and $k$ is the order of operation.

Performing a similar substitution (\ref{am13}) in (\ref{am2}),  after some 
 transformations we obtain  the following  equations:
\begin{equation} \label{am19}
\dot{\theta}_p =\sum_{s=1}^n \left\{\theta_p,\theta_s\right\}\frac{\partial B(\theta,\tau)}{\partial\theta_s},\ \ (p=1,\ldots.n)
\end{equation}
or
\begin{equation} \label{am20}
\dot{\theta}_p =\sum_{s=1}^n a_{ps}(\theta)\frac{\partial B(\theta,\tau)}{\partial\theta_s},\ \ (p=1,\ldots.n),
\end{equation}
where the function $B(\theta,\tau)$ is the Hamiltonian expressed in terms of the variables $\theta$.
The system of ordinary differential equations (\ref{am19}) coincides with the variational Lagrange equations
in the general form \cite{Smart}.
Therefore, we can interpret our special variables, which can now include the Keplerian orbital elements.

If we then  compare the right-hand sides of (\ref{am16}) and (\ref{am19}), we see that the form of the  expressions 
\begin{equation} \label{am21}
\sum_{s=1}^{n} \left\{\theta_{p},\theta_{s} \right\}\frac{\partial W_{0k}(\theta)}{\partial \theta_s}\;\                                                   
\text{and}\ \ \sum_{s=1}^n \left\{\theta_p,\theta_s\right\}\frac{\partial B(\theta,\tau)}{\partial\theta_s}
\end{equation}
coincides.
Thus, the right-hand sides of equations (\ref{am19}) or (\ref{am20}) can be used to form the matrix $\Psi$, if we know
the generating function.
Consequently, equations (\ref {am19}) and (\ref{am20}) can be adopted as basic equations;  then, using the Lie generator 
(\ref{am18}), all calculations  can be performed in the variables $\theta$ remaining within the framework of the
 Lie transformation  theory for canonical systems.   In the next section, we define the meaning of the obtained relationships.

\subsection{Birkhoff's equations}
\label{sec:2.3} 
 Consider an extended phase space with $m=2l+1$ dimensions, where $l$ is the number of degrees of freedom of the system.
 Let us turn to the usual notations of canonical variables: $q, p$.  In the space under consideration, we then have the following
 set of generalized coordinates and generalized momenta:  $q\:(q_1,\ldots,q_l); p\:(p_1,\ldots,p_l)$. The additional
 variable  is the time  $t$. Let the function $H (q, p,t)$ be the Hamiltonian of a dynamical system.  For this system, we introduce  the following 1-form (or Pfaffian):
\begin{equation} \label{am22}
\omega^1=p\:dq-H\:dt.
\end{equation}
It is known \cite{Arnold} that in the extended phase space, the phase trajectories of a dynamical system with the Hamiltonian
$H (q,p,t)$ are the vortex lines of the form $\omega^1$. If the  variables used are non-canonical,
then the Pfaffian (\ref{am22}) can be written in a large number of different forms \cite{Broucke}. This means that we can write
the equations of motion in any suitable system of coordinates  in the extended phase space with $m$ dimensions.
It follows, therefore, the ratio \cite{Arnold}:
\begin{equation} \label{am23}
p\:dq-H\:dt=\Theta_1\:d\theta_1+\ldots+ \Theta_{m}\:d\theta_{m},
\end{equation}
 where $\Theta_i=\Theta_i(\theta)$ are smooth functions, and $\theta_{m}=t,\Theta_{m}=H(\theta)$. In addition, 
 the function $H(\theta)$ will be denoted by $B$. 

In the variables  $\theta$, the Pfaffian has the form 
\begin{equation} \label{am24}
\Phi^1=\Theta_1\:d\theta_1+\ldots+ \Theta_{m}\:\theta_{m}.
\end{equation}
The set
\begin{equation} \label{am25}
\bm {P}=(\Theta_1,\ldots,\Theta_{m})
\end{equation}
is called the Pfaff vector of the considered dynamic system \cite{Broucke}.
If the Pfaffian (\ref{am24}) is an exact differential, then the curl of  $\bm {P}$ is equal to zero.
From here, we follow the dynamic equations: 
\begin{equation} \label{am26}
\sum^{m}_{j=1}\left(\frac{\partial \Theta_{i}}{\partial \theta_{j}}-\frac{\partial \Theta_{j}}{\partial \theta_{i}}\right)d\theta_{j}=\sum^{m}_{j=1}b_{ij}d\theta_{j}=0, \ \ \ i= 1,\ldots, m.
\end{equation}
These equations are considered in the work of \cite{Broucke}. 

The Eq. (\ref{am26}) is  represented  in the   phase space with $m=2l+1$ dimensions.  However,  
further study of these equations is not convenient, since $b_{ij}$ is the  askew-symmetric matrix   of
odd order, and its determinant  is equal to zero.

We can  obtain the dynamic equations in the even-dimensional phase space  using  the Pfaff-Birkhoff variational principle.
Let us define 1-form in phase space with $n=2l$ dimensions 
\begin{equation} \label{am27}
\varpi^1=\sum^{n}_{\nu=1} R_\nu(\theta)\:d\:\theta_\nu =\sum^{n}_{i=1}\Theta_{i}\:d\:\theta_i ,
\end{equation}
where $R_\nu$ is traditionally accepted designation of  the Birkhoff functions   (see e.g., \cite{Santilli}). In this
work, we assume that a set of  Birkhoff's functions $(R_1,\ldots,R_n)$  coincides with the
elements $(\Theta_1,\ldots,\Theta_{(m-1)})$  of the  Pfaff vector.
In addition,  for convenience of presentation,
we denote the Birkhoff  functions by a set: $(\Theta_1,\ldots,\Theta_n)$.

In the introduced notations, the Pfaff-Birkhoff variational principle   can be written as follows:
\begin{equation} \label{am28}
\delta\:S=\delta\:\int_{t_1}^{t_2}\left(\sum^{n}_{j=1}\Theta_j\:\dot{\theta_j}-B\right)=0.
\end{equation}
The integrand in (\ref{am28}) is a linear function of the derivatives $\dot{\theta}_j$ with coefficients 
$\Theta_j(\theta),\;B(\theta)$.  By using the first-order
 variation $\delta\:S=0$
with fixed end-point conditions $\delta \theta_j(t_1)=0,\ \delta \theta_j(t_2)=0,\;j=1,\ldots,n $, the autonomous Birkhoff equations 
can be obtained as follows \cite{Santilli}:
 \begin{eqnarray} \label{am29}
\sum^{n}_{p=1}\left(\frac{\partial \Theta_{s}}{\partial \theta_{p}}-\frac{\partial \Theta_{p}}{\partial \theta_{s}}\right) \dot{\theta_{p}}-\frac{\partial B}{\partial \theta_s}=0,\ \ (s=1,\ldots,n;\:n=2l).
\end{eqnarray}
In (\ref{am29}), the function $B(\theta)$ is called the Birkhoffian, and
\begin{equation} \label{am30}
\Omega_{ps}=\frac{\partial \Theta_{s}}{\partial \theta_{p}}-\frac{\partial \Theta_{p}}{\partial \theta_{s}}
\end{equation}
is called the Birkhoff tensor. Using the terminology of \cite{Santilli}, it is a covariant Birkhoff tensor.  

The symplectic structure $\Omega$ in the space under consideration is defined as the external
 differential of the 1-form $\varpi^1$
\begin{eqnarray} \label{am31}
\Omega=d\:\varpi^1
\end{eqnarray}
or
\begin{equation} \label{am32}
\Omega=\sum_{p=1}^n\Omega_{ps}\:d\theta_s\:\wedge\:d\theta_p,\ \ (s=1,\ldots,n),
\end{equation}
where $\Omega_{ps}$ is defined by the ratio (\ref{am30}).

It can be shown that the matrix $(\Omega_{ps})$ coincides with the skew-symmetric matrix $(\omega_{ps})$ of the Lagrange brackets.
From (\ref{am23}) and (\ref{am11}) we find
\begin{equation} \label{am33}
\Theta_s=p_i\frac{\partial{q_i}}{\partial{\theta_s}}=\eta_{l+i}\frac{\partial{\eta_i}}{\partial{\theta_s}}.
\end{equation}
Next, we substitute
\begin{equation} \label{am34}
\Theta_s=\eta_{l+i}\frac{\partial{\eta_i}}{\partial{\theta_s}},\ \ \Theta_p=\eta_{l+i}\frac{\partial{\eta_i}}{\partial{\theta_p}}
\end{equation}
to $(\Omega_{ps})$. The result is a skew-symmetric square matrix, each element of which represents the Lagrange bracket
\begin{equation} \label{am35}
\omega_{ps}=\sum_{i=1}^l\left(\frac{\partial{\eta_{l+i}}}{\partial{\theta_p}}\frac{\partial{\eta_i}}{\partial{\theta_s}}-
\frac{\partial{\eta_{l+i}}}{\partial{\theta_s}}\frac{\partial{\eta_i}}{\partial{\theta_p}}\right).
\end{equation}
We suppose that the dynamic system is non-singular, i.e.,
\begin{equation} \label{am36}
det(\Omega_{ps})\neq{0}.
\end{equation}
Thus, matrix $(\omega_{ps})$ is   non-degenerate.  Therefore there is an inverse matrix
\begin{equation} \label{am37}
(a_{ps})= (\omega_{ps})^{-1}.
\end{equation}
Matrix (\ref{am37}) is also a skew-symmetric square matrix, and each element of this matrix is the Poisson bracket.
A detailed derivation can be found in \cite{Smart}.
The matrix $(a_{ps})$ consists of the  Poisson brackets of the following form:
 \begin{equation} \label{am38}
a_{ps}=\sum_{i=1}^l\left(\frac{\partial{\theta_{p}}}{\partial{\eta_{l+i}}}\frac{\partial{\theta_s}}{\partial{\eta_i}}-
\frac{\partial{\theta_{p}}}{\partial{\eta_i}}\frac{\partial{\theta_s}}{\partial{\eta_{l+i}}}\right).
\end{equation}

 From the above, it follows that the matrix $(a_{ps})$   defines the
 Birkhoff tensor, which is expressed in terms of Poisson brackets. 
In accordance with the terminology of
 \cite{Santilli},  $a_{ps}$  is the contravariant  Birkhoff tensor, also called the Lie tensor. Thus, Eq. (\ref{am20})
 are the Birkhoff equations, which are presented in a contravariant form.

In this  section,  we have shown that  the representation of the Lie generator  
for canonical systems in terms of the special variables of the phase space leads to its expression in terms of 
the Birkhoff tensor $a_{ps}$. It follows that all the components of the algorithm can be expressed using tensors. 
This makes it possible to use of the autonomous Birkhoff's equations  for solving of  problems  of dynamics
in the non-canonical variables in the phase space.
 
\section{Some applications}
\label{sec3}


\subsection{Closed form representation of the short-period perturbations in the motion of a satellite. The first example}
\label{sec3.1}

We  now consider the motion of a satellite under the action of gravity of the planet and the Sun, provided that all three
bodies are material points. The satellite has an infinitely small mass, i.e., it has no gravitational effect on the other 
two bodies. The central body is a planet with mass $m_0 $. The Sun, with mass $m'$, moves around the planet in a circular
 orbit located in the main coordinate plane. We use a rotating coordinate system that is the same as that used by  Hill's
group in their studies of lunar theory. In  the Delaunay canonical  elements,  $p=(L,G,H)$, $q=(l,g,h)$, the equations
 of motion have the form
\begin{equation} \label{am39}
\dot{q}_k=\frac{\partial{F}}{\partial{p}_k},\ \ \ \ \dot{p}_k=-\frac{\partial{F}}{\partial{q}_k},\ \ \ k=1,2,3,
\end{equation}
where $F(p,q)$ is the Hamiltonian of the perturbed problem:
\begin{equation} \label{am40}
F=F_{00}+\nu\:F_{01}+\nu^2\,F_{02},
\end{equation}

\begin{equation} \label{am41}
F_{00}=-\frac{\mu^2}{2L^2},\ \ \ \ \  F_{01}=-H,
\end{equation}
\begin{equation} \label{am42}
F_{02}=-\sum_{n=2}^\infty\:r^n\ P_n\![\cos{(S)}].\ \ \ \
\end{equation}
In (\ref{am40}), (\ref{am41}) and  (\ref{am42})  $\mu=G\:m_0 $; $G$ is the constant of universal gravitation; $P_n$ are the
Legendre polynomials (in this example, assume n=2), $\nu=n'$ is the small parameter, and $n'$ is the mean motion of the Sun.
It is a constant formal parameter.
The strength of the disturbances is characterized by the implicit parameter $m=n'/n$, where $n$ is the mean motion of
the satellite, and $H$ is  the Delaunay variable. In  (\ref{am41}) and  (\ref{am42}) it is  taken into account  that the canonical
elements $p=(L,G,H)$ are variables of the action.

In the ratio (\ref{am42})
\begin{equation} \label{am43}
\cos{(S)}=\cos{(f+g)}\:\cos{h} - \sin{(f+g)}\:\sin{h}\:\cos{i},
\end{equation}
where $f$ is the true anomaly of the satellite; $g=\omega$, $\omega$ determines the argument of periapsis of the satellite orbit;
$h=\Omega-\lambda'$,  $\Omega$ is the longitude of the ascending node of the satellite orbit, $\lambda'$ is the mean longitude of
the Sun;
$i$ is the inclination of the satellite orbit to the primary coordinate plane.

The first term $F_{00}$ in (\ref{am40}) is caused by the attraction of the planet in the absence of perturbations,
 $\nu\:F_{01}$ is a term that appears due to the use of a rotating coordinate system, and the third term $\nu^2\,F_{02}$
is the disturbing function.

 We consider the non-resonant case, and  suppose that Hamiltonian is an analytic function of all variables and 
has period $2\pi$ in all angular variables. The Lee transformation method, in combination with the  averaging 
of the disturbing function over the mean anomaly $l$ of the  satellite, was chosen in oder to eliminate terms of a  short
period   from   the Hamiltonian. The Pfaffian of this problem is given by
\begin{equation} \label{am44}
\Phi=L\:dl+G\:dg+H\:dh-F\:dt.
\end{equation}

 In this example,  instead of the mean anomaly $l$, we use   eccentric anomaly $u$ of the satellite in all  expressions,
 in order to  avoid an expansion  in  powers of  eccentricity $e$ of the satellite orbit. All  analytical
expansions are carried out as the truncated series in  $ m=n'/n $ , but the coefficients of these series are in closed form.
 
We now introduce the  variables  $\epsilon=(\alpha,\:\eta,\:\gamma,\:u,\:g,\:h)$
using the  ratios:
\begin{eqnarray} \label{am45}
L=\alpha\:\mu_1,\ \ G=\alpha\:\mu_1\eta,\ \ H=\alpha\:\mu_1\eta\:\gamma,\ 
l=u-\sqrt{1-\eta^2}\sin{u},\ \ g=g,\ \ h=h,\ 
\end{eqnarray}
where $\mu_1=\sqrt{\mu}$; $\alpha=\sqrt{a}$,  $a$ is the semi-major axis of the satellite orbit, $\eta=\sqrt{1-e^2}$, and 
$\gamma=\cos{i}$. The Delaunay elements are presented in the form (\ref{am45}) for convenience of computation using analytical 
computer systems.

Pfaffian in the terms of the variables $\epsilon$ has the form:
\begin{eqnarray} \label{am46}
\Phi =\alpha\:\mu_1\:\frac{\eta}{\sqrt{1-\eta^2}}\:\sin{u}\:d\eta+\alpha\:\mu_1\:(1-\sqrt{1-\eta^2}\:\cos{u})\:du +\nonumber\\
\alpha\:\mu_1\eta\:dg +\alpha\:\mu_1\eta\:\gamma\:dh-B(\epsilon)\:d\:t.
\end{eqnarray}
The coefficients of the differentials in  (\ref{am46}) define the Pfaff vector $\bm P$ (\ref{am25}) of the system.
The Birkhoff functions, as we previously identified, are included in the Pfaff vector, and are given  by the following set:
 
\begin{equation} \label{am47}
\left(0,\:\alpha\:\mu_1\:\frac{\eta}{\sqrt{1-\eta^2}}\sin{u},\:0,\;\alpha\:\mu_1\:(1-\sqrt{1-\eta^2}\:\cos{u}),\;\alpha\:\mu_1\eta,\;\alpha\:\mu_1\eta\:\gamma\right).
\end{equation}

Next, we find  the equations of motion by making use of (\ref{am20}) ( Eq. (\ref{am29}) can be used to 
verify the correctness of the output ):
\begin{eqnarray} \label{am48}
\frac{d\alpha}{dt}&=&-\frac{\alpha^{2}}{\mu_1 r}\frac{\partial{B}}{\partial{u}},\nonumber\\
\frac{d\eta}{dt}&=&\frac{\alpha\,\eta}{\mu_1 r}\frac{\partial{B}}{\partial{u}}-\frac{1}{\mu_1\alpha}\frac{\partial{B}}{\partial{g}},\nonumber\\
\frac{d\gamma}{dt}&=&\frac{\gamma}{\mu_1\alpha\:\eta}\frac{\partial{B}}{\partial{g}}-\frac{1}{\mu_1\alpha\:\eta}\frac{\partial{B}}{\partial{h}},\nonumber\\
\frac{du}{dt}&=&\frac{\alpha^{2}}{\mu_1 r}\frac{\partial{B}}{\partial{\alpha}}-\frac{\alpha\,\eta}{\mu_1 r}\frac{\partial{B}}{\partial{\eta}}+\frac{\alpha\eta \sin{u}}{\mu_1r\:\sqrt{1-\eta^{2}}}\frac{\partial{B}}
{\partial{g}},\nonumber\\
\frac{dg}{dt}&=&\frac{1}{\mu_1\alpha}\frac{\partial{B}}{\partial{\eta}}-\frac{\alpha\eta \sin{u}}{\mu_1r\sqrt{1-\eta^{2}}}\frac{\partial{B}}{\partial{u}}-\frac{\gamma}{\mu_1\alpha \:\eta}\frac{\partial{B}}{\partial{\gamma}},\nonumber\\
\frac{dh}{dt}&=&\frac{1}{\mu_1\alpha  \eta}\frac{\partial{B}}{\partial{\gamma}},
\end{eqnarray}
where $B(\epsilon)$ is the Birkhoffian ( Hamiltonian expressed in terms of the variables $\epsilon$):
\begin{equation} \label{am49}
B=B_{00}+\nu\:B_{01}+\nu^2\,B_{02},
\end{equation}
\begin{equation} \label{am50}
B_{00}=-\frac{\mu_{1}^2}{2\alpha^2},\ \ \ \ \  B_{01}=-\mu_1\alpha\eta\gamma.
\end{equation}
In (\ref{am49}), $B_{02}$ is the function (\ref{am42}) expressed in terms of the variables $\epsilon$. 
As shown by \cite{Birkhoff}, in the autonomous case the Birkhoffian is 
the  integral of motion.

The Birkhoff equations (\ref{am48}) can be written in general form as follows: 
\begin{equation} \label{am51}
\dot{\epsilon}_i=\sum_{j=1}^{6} a_{i\:j}(\epsilon)\frac{\partial{B(\epsilon)}}{\partial\epsilon_j},\ \ i=1,2.\ldots,6,
\end{equation}
where $a_{i\:j}(\epsilon)$ is a skew-symmetric matrix, obtained from (\ref{am38}):
\[\left(\begin{array}{lccccr}
\medskip
0                       &  0                            &  0                                 & -\frac{\alpha^2}{\mu_1\:r}                                   &  0                                                     &  0 \\
\medskip
0                       &  0                            &  0                                 &  \frac{\alpha\:\eta}{\mu_1\:r}                             & -\frac{1}{\mu_1\:\alpha}                                    &  0\\
\medskip
0                       &  0                            &  0                                 &  0                                                      &  \frac{\gamma}{\mu_1\:\alpha\:\eta}                       &  -\frac{1}{\mu_1\:\alpha\:\eta}\\
\medskip
\frac{\alpha^2}{\mu_1\:r}  & -\frac{\alpha\:\eta}{\mu_1\:r}    &  0                                 &  0                                                      &  \frac{\alpha\:\eta\:\sin{u}}{\mu_1\:r\:\sqrt{1-\eta^2}}  &  0 \\
\medskip
0                       &  \frac{1}{\mu_1\:\alpha}         & -\frac{\gamma}{\mu_1\alpha\:\eta}    & -\frac{\alpha\:\eta\:\sin{u}}{\mu_1r\:\sqrt{1-\eta^2}}     &  0                                                     &  0 \\
\medskip
0                       &  0                            &  \frac{1}{\mu_1\:\alpha\:\eta}        &  0                                                      & 0                                                      &  0
\end{array}\right)\]

In addition, we introduce the generating function in the form:
\begin{equation}\label{am52}
W({\epsilon})= \nu W_{01}(\epsilon)+\nu^2 W_{02}(\epsilon)+\dots+\nu^k W_{0k}(\epsilon)+\ldots.
\end{equation}
The Lie generator   is represented as follows(\ref{am18}):
\begin{equation} \label{am53}
L_{k}f_{ij}(\epsilon)=\sum_{p=1}^6 W_{pk}\frac{\partial{f_{ij}(\epsilon)} }{\partial \epsilon_p},
\end{equation}
where $f_{ij}(\epsilon)$ are some analytical functions, $ W_{pk} $ are elements of the matrix 
$\Psi=\left(W_{pk}\right), (p=1,\ldots,6;k=1,\ldots,5)$, and
k  is the order of transformation. Each column of the matrix $\Psi$ is determined by the ratio:
\begin{equation} \label{am54}
W_{ik}=\sum_{j=1}^6 a_{ij}(\epsilon)\frac{\partial\:W_{0k}}{\partial\:\epsilon_j},\ \ \ (i=1,\ldots,6).
\end{equation}

As an example, let us write the expression (\ref{am54}) explicitly for k-th order
\begin{eqnarray} \label{am55}
W_{1k}&=&-\frac{\alpha^{2}}{\mu_1 r}\frac{\partial{W_{0k}}}{\partial{u}},\nonumber\\
W_{2k}&=&\frac{\alpha\,\eta}{\mu_1 r}\frac{\partial{W_{0k}}}{\partial{u}}-\frac{1}{\mu_1\alpha}\frac{\partial{W_{0k}}}{\partial{g}},\nonumber\\
W_{3k}&=&\frac{\gamma}{\mu_1\alpha\:\eta}\frac{\partial{W_{0k}}}{\partial{g}}-\frac{1}{\mu_1\alpha\:\eta}\frac{\partial{W_{0k}}}{\partial{h}},\nonumber\\
W_{4k}&=&\frac{\alpha^{2}}{\mu_1 r}\frac{\partial{W_{0k}}}{\partial{\alpha}}-\frac{\alpha\,\eta}{\mu_1 r}\frac{\partial{W_{0k}}}{\partial{\eta}}+\frac{\alpha\eta \sin{u}}{\mu_1r\:\sqrt{1-\eta^{2}}}\frac{\partial{W_{0k}}}
{\partial{g}},\nonumber\\
W_{5k}&=&\frac{1}{\mu_1\alpha}\frac{\partial{W_{0k}}}{\partial{\eta}}-\frac{\alpha\eta \sin{u}}{\mu_1r\sqrt{1-\eta^{2}}}\frac{\partial{W_{0k}}}{\partial{u}}-\frac{\gamma}{\mu_1\alpha \:\eta}\frac{\partial{W_{0k}}}{\partial{\gamma}},\nonumber\\
W_{6k}&=&\frac{1}{\mu_1\alpha  \eta}\frac{\partial{W_{0k}}}{\partial{\gamma}}.
\end{eqnarray}
Here, the functions $W_{0k}$  are found  from the homological equation \cite{Ferraz-Mello}
\begin{equation} \label{am56}
L_{k}B_{00}=B_{0k}^*-\Delta'_{k}.
\end{equation}  
The expression for $\Delta'_{k}$ will be determined later.
The function $B_{0k}^*$  is found from the relation
\begin{equation} \label{am57}
B_{0k}^*=\frac{1}{2\pi}\int_{0}^{2\pi}\Delta'_{k}dl=\frac{1}{2\pi}\int_{0}^{2\pi}\Delta'_{k}(1-\sqrt{1-\eta^2}\cos(u))du.
\end{equation}

For a better understanding of the algorithm let us introduce the Lie triangle (\ref{am9}) in the form: 
\begin{eqnarray}
\label{am58}
 & B_{00} & B_{01} \;\;  B_{02} \;\; B_{03} \;\; B_{04}  \ldots \nonumber \\
 & B_{10} & B_{11} \;\;  B_{12} \;\;  B_{13} \ldots \nonumber \\
 & B_{20} & B_{21} \;\;  B_{22} \ldots \nonumber \\
 & B_{30} & B_{31} \;\; \ldots \nonumber \\
 & B_{40} &\ \ldots
 \end{eqnarray}
In (\ref{am58}),  functions  $B_{00}, B_{01}$, and $B_{02}$ are the coefficients of the original Birkhoffian (\ref{am49}), 
and $B_{03}=B_{04}=B_{05}=\ldots=0$.

The Birkhoffian of  the transformed system is written as 
\begin{equation} \label{am59}
B^*=B_{00}^*+\nu\:B_{01}^*+\nu^2\,B_{02}^*+\nu^3\,B_{03}^*+\nu^4\,B_{04}^*+\nu^5\,B_{05}^*+\ldots.
\end{equation}
In the algorithm process, short-period terms appear at higher orders; therefore, we also  find 
the functions $B_{03}^*,\;B_{04}^*,\;B_{05}^*,\ldots$

The transformation  $B(\epsilon)\rightarrow B^*(\epsilon)$  can be carried out using the following recursive algorithm:
\begin{eqnarray}
\label{am60}
\Delta_{k}&=&\sum_{i=0}^k B_{i,k-i},\ k=0,\dots,m;\nonumber\\
B_{i,k-i}& =&\frac{1}{i}\sum_{\rho=0}^{k-i} L_{\rho+1}B_{i-1,k-1-\rho}, \ i\neq{0}.
\end{eqnarray}

The function $\Delta'_{k}$ is determined from the relation (\ref{am56}) as follows: 
 \begin{equation} \label{am61}
\Delta'_{k}=\Delta_{k}-L_{k}B_{00}.
\end{equation}

 Let us consider the first  three normalization steps in  detail.
 For the most compact form of writing, we assume $\gamma=1$, $h=0$, $g=\omega-\lambda'$ and use the ratio $e^2+\eta^2 =1$.
That is, we consider a planar version of the problem.

 Oder 0: $B_{00}=B_{00}^*=\displaystyle{-\frac{\mu_1^2}{2\alpha^2}}$. 

Oder 1: If $B_{02}$ is neglected in  (\ref{am49}), Eq. (\ref{am48}) are integrable, and the transformation is 
identical. Therefore, $W_{01}=0$. All elements of the first column of the matrix $\Psi$ are equal to zero and
$\nu B_{01}^*= -\nu \mu_1\alpha\eta$.

 Oder 2: At this point, we find the diagonal elements of the second order of the Lie triangle, and we then define 
the homological equation:
\[
B_{20}=0,\ \ B_{11}=L_2 B_{00},\ \  \Delta'_{2}=B_{20}+B_{02}=B_{02},
\]
\begin{equation} \label{am62}
L_{2}B_{00}=B_{02}^*-B_{02}.
\end{equation}

Let us  enter the expression for $B_{02}$. In the assumptions adopted above, the function $B_{02}$ is written as follows:
\begin{equation} \label{am63}
B_{02}=r^2\left(\frac{1}{2}-\frac{3}{2} \cos^2 S\right),\ \  \cos S= \cos(f+g).
\end{equation}
Then, using the formulas of the theory of Keplerian motion
\[
r=\alpha^2(1-e\cos u),\ \ \sin f=\frac{\alpha^2}{r}\eta \sin u,\ \  \cos f=\frac{\alpha^2}{r}(\cos u-e),
\]
where $u$ is the eccentric anomaly, we obtain the expression for the disturbing function $\nu^2 B_{02}$ in the form:
\begin{eqnarray} \label{am64}
\nu^2 B_{02}=&\frac{1}{16}\nu^2 \alpha^4 (4+2 e^2+18e^2 \cos{2 g}-3(-2+e^2+2 \eta)\cos(2g-2u)+ \nonumber \\
&(-12e+12e\eta)\cos(2g-u)-8e \cos u + 2e^2\cos 2u+\nonumber \\
&(6-3e^2+6\eta)\cos(2g+2u)+(-12e-12e\eta)\cos(2g+u)).
\end{eqnarray}
Averaged over  mean anomaly $l$ of the satellite, the function $B_{02}^{*}$ is found from the relation 
\begin{equation} \label{am65}
B_{02}^*=\frac{1}{2\pi}\int_{0}^{2\pi}B_{02}dl=\frac{1}{2\pi}\int_{0}^{2\pi}B_{02}(1-e\cos(u))du.
\end{equation}
Finally, we have
\[
\nu^2 B_{02}^*=\frac{\nu^2 \alpha^4}{4}\left(1-\frac{3 }{2}e^2+\frac{15}{2}e^2 \cos(2g)\right)=
\]
\[
\frac{\nu^2}{n^2}\frac{\mu_1^2}{\alpha^2}\frac{1}{4}\left(1-\frac{3 }{2}e^2+\frac{15}{2}e^2 \cos(2g)\right).
\]
Here, the   multiplier $\mu_1^2/\alpha^2$  has the dimension of energy.

We then find 
\[
B_{00}= -\frac{\mu_1^2}{2\alpha^2},\ \  L_{2}B_{00}=-\frac{\alpha^{2}}{\mu_1 r}\frac{\partial{W_{02}}}{\partial{u}}\frac{\partial{B_{02}}}{\partial{\alpha}}=-\frac{\mu_1}{\alpha r}\frac{\partial{W_{02}}}{\partial{u}}. 
\]
Now, the homological equation is written as
\begin{equation} \label{am66}
\frac{\mu_1}{\alpha r}\frac{\partial{W_{02}}}{\partial{u}}= B_{02}-B_{02}^*
\end{equation}
and
\begin{equation} \label{am67}
W_{02}=\int \frac{\alpha^3}{\mu_1}(B_{02}-B_{02}^*)(1-e\cos u)du.
\end{equation}
The  function $W_{02}$ should be periodic in mean anomaly $l$. Therefore, we find
\[
C=\frac{1}{2\pi}\int_{0}^{2\pi} W_{02}(1-e\cos u)du=\frac{15}{16 \mu_1}\alpha^7 e^2 \eta\sin(2g),\ \ W_{02}=W_{02}-C.
\]
The final expression for the function $\nu^2 W_{02}$ has the form
\begin{equation} \label{am68}
\nu^2 W_{02}=\nu^2\frac{\alpha^7}{\mu_1}W'_{02}=\left(\frac{\nu^2L}{n^2}\right)\; W'_{02},
\end{equation}
where $\displaystyle {n=\frac{\mu_1}{\alpha^3}}$ $-$ mean motion of the satellite, $L=\mu_1 \alpha$ $-$ is an element of Delaunay, 
 which determines  the dimension  of the generating function,
\begin{eqnarray} \label{am69}
W'_{02}& = &\frac{1}{32}(-30e^2 \eta\sin(2 g)+(2e-e^3-2e\eta)\sin(2 g-3u)+\nonumber \\
       &   &(-6-3e^2+6\eta+6e^2\eta)\sin(2 g-2u)+(-16e+6e^3)\sin u\nonumber \\
       &   &(30e-15e^3-30e\eta)\sin(2 g-u)+6e^2 \sin(2u)-\frac{2}{3}e^3\sin(3u)+\nonumber \\
       &   &(-30 e+15 e^3 -30 e \eta)\sin(2g+u)+\nonumber \\
       &   &(6+3e^2+6\eta+6e^2\eta)\sin(2g+2u)+\nonumber \\
			 &   &(-2e+e^3-2e\eta)\sin(2g+3u)).
\end{eqnarray}
In accordance with the assumptions made above for the planar version of the problem, we define the second column 
of the matrix $\Psi$:
\begin{eqnarray} \label{am70}
W_{12}&=&-\frac{\alpha^{2}}{\mu_1 r}\frac{\partial{W_{02}}}{\partial{u}},\nonumber\\
W_{22}&=&\frac{\alpha\,\eta}{\mu_1 r}\frac{\partial{W_{02}}}{\partial{u}}-\frac{1}{\mu_1\alpha}\frac{\partial{W_{02}}}{\partial{g}},\nonumber\\
W_{32}&=&\frac{\alpha^{2}}{\mu_1 r}\frac{\partial{W_{02}}}{\partial{\alpha}}-\frac{\alpha\,\eta}{\mu_1 r}\left(\frac{\partial{W_{02}}}{\partial{\eta}}-\frac{\eta}{e}\frac{\partial{W_{02}}}{\partial{e}}\right)+\frac{\alpha\eta \sin{u}}{\mu_1r\:e}\frac{\partial{W_{02}}}
{\partial{g}},\nonumber\\
W_{42}&=&\frac{1}{\mu_1\alpha}\left(\frac{\partial{W_{02}}}{\partial{\eta}}-\frac{\eta}{e}\frac{\partial{W_{02}}}{\partial{e}}\right)-\frac{\alpha\eta \sin{u}}{\mu_1r\: e}\frac{\partial{W_{02}}}{\partial{u}}.
\end{eqnarray}
The last step of this phase is to compute the function $B_{11}$
\[
B_{11}= L_2 B_{00}=-\frac{\alpha^2}{\mu_1r}\frac{\partial W_{02}}{\partial u}\frac{\partial B_{00}}{\partial \alpha},
\]
\begin{eqnarray} \label{am71}
\nu^2 B_{11}=&\displaystyle{\frac{1}{16}\frac{\nu^2}{n^2}\frac{\mu_1^2}{\alpha^2}} (- e^2-12e^2 \cos{2 g}+3(-2+e^2+2 \eta)\cos(2g-2u)+ \nonumber \\
&(-12e+12e\eta)\cos(2g-u)-8e \cos u + 2e^2\cos 2u+\nonumber \\
&3(2-3e^2+2\eta)\cos(2g+2u)-12(e+e\eta)\cos(2g+u)).
\end{eqnarray}

 Oder 3: In the third oder, similar calculations   lead to the following results:
\[
\Delta'_3=B_{30}+B_{21}+B_{12}+B_{03}-L_3\;B_{00};\ \  B_{30}=0,\ B_{21}=0,\ B_{03}=0;
\]
\[
\Delta'_3=B'_{12}=L_2\;B_{01}=L_2\;(-\mu_1\alpha\eta)=-\frac{\alpha^2}{\mu_1\:r}\frac{\partial W_{02}}{\partial u}\frac{\partial B_{01}}{\partial \alpha}+
\]
\[
\left(\frac{\alpha\eta}{\mu_1\:r}\frac{\partial W_{02}}{\partial u}-\frac{1}{\mu_1\alpha}\frac{\partial W_{02}}{\partial g}\right)\frac{\partial B_{01}}{\partial \eta}=\frac{\partial W_{02}}{\partial g};
\]
\[
B_{03}^*=\frac{1}{2\pi}\int_{0}^{2\pi}\Delta'_3 (1-e\cos(u))du = 0;\ \ \ \ \ L_3\;B_{03}=B_{02}^*-\Delta'; 
\]
\[
\frac{\mu_1}{\alpha r}\frac{\partial W_{03}}{\partial u}=\frac{\partial W_{02}}{\partial g};\ \ \ \ \  W_{03}=\int \frac{\alpha^3}{\mu_1}(1-e\cos u)\Delta'_3 du.
\]
We then  define $W_{03}$ in the form
\begin{eqnarray} \label{am72}
 \nu^3 W_{03}&=&\frac{\nu^3\;L}{384 n^3}W'_{03},\nonumber \\
W'_{03}& = &-36e^2(-22+9e^2) \sin(2 g)+\nonumber \\
       &   & 4e(10(-1+\eta)+e^2)(-1+6\eta))\sin(2 g-3u)+\nonumber \\
       &   &(72+228-96e^4-72\eta-264e^2\eta)\sin(2 g-2u)+\nonumber \\
       &   &(6e^2-3e^4-6e^2\eta)\sin(2 g-4u)+\nonumber \\
       &   &(-792 e+324 e^3+792e\eta -288 e^3 \eta)\sin(2g-u)+\nonumber \\
       &   &(72+228e^2-96e^4+72\eta+264e^2\eta)\sin(2g+2u)+\nonumber \\
			 &   &(-792 e+324 e^3-792e\eta +288 e^3 \eta)\sin(2g+u)+\nonumber \\
			 &   &(-40e-4e^3-40e\eta-24e^3\eta)\sin(2g+3u)+\nonumber \\
			 &   &( 6e^2-3e^4+6e^2\eta)\sin(2g+4u).
\end{eqnarray}
	Next, we define the third column of the matrix $\psi$ by the scheme (\ref{am70}). 
		 The last step of this phase is to compute the function $B_{12}$:
\[
B_{12}= L_2 B_{01}+L_{3} B_{00}=0.			
\]

Let us briefly discuss the calculations of the fourth order. 

Function $\Delta'_4$ is defined as follows:
\[
\Delta'_4=B'_{13}+B_{22}; \ B'_{13} =L_2 B_{02}+L_3 B_{01};\ B_{22}=L_2 B_{11}.
\]
For example, consider the operation $L_2 B_{11}$:
\begin{equation} \label{am73}
L_{2}B_{11}(\epsilon)=\sum_{p=1}^4 W_{p2}\frac{\partial{B_{11}(\epsilon)}}{\partial \epsilon_p}.
\end{equation}
In the process of computation of  the derivatives   $\displaystyle{\frac{\partial{B_{11}(\epsilon)}}{\partial \epsilon_p}}$ 
we must keep in mind  that $e^2+\eta^2=1$.

Evaluating expressions in the form 
\begin{equation} \label{am74}
W_{p2} \frac{\partial{B_{11}(\epsilon)}}{\partial \epsilon_p}
\end{equation}  
 is rather cumbersome, but the algorithm is designed  to minimize the number of such operations. It is easy to see that
 the expression (\ref{am73}) can be represented 
in the form of the generalized Poisson brackets (\ref{am16}), but in this case, the number of operations of type (\ref{am74}) 
is increased about twofold. Thus, the use of the matrix $\Psi$ minimizes the number of cumbersome operations.

We see from the relations (\ref{am70}) that the functions  $W_{ik}$ have a multiplier $r^{-1}$. 
However, because in the process of averaging  over  the mean anomaly we have
\[
\left\langle \frac{\alpha^2}{r}\right\rangle=1,\ \ \left\langle\frac{\alpha^2}{r}\sin ju \right\rangle=0,\ j\geq1,
\]
and expressions of the form (\ref{am67}) are integrated by using formula 
\[
d\:l=\frac{r}{\alpha^2}du,
\]
negative powers of $r$ do not appear in the final results.

The function $\nu^4 B_{04}^*$ has the form:
\begin{equation} \label{am75}
\nu^4 B_{04}^*= \nu^4\frac{\alpha^{10}}{\mu_1^2} B_{04}^{'*}=\frac{\nu^4}{n^4}\frac{\mu_1^2}{\alpha^2} B_{04}^{'*},
\end{equation}  
\[
B_{04}^{'*}=\frac{1}{16}\left(\frac{49}{4}-\frac{873}{4}e^2+\frac{4347}{32}e^4-\left(\frac{333}{4}e^2-\frac{237}{8}e^4\right)\cos 2g+\frac{615}{32}\cos 4g\right),
\]
and for the generating function we have
\begin{equation} \label{am76}
\nu^4 W_{04}=\frac{\nu^4\;L}{ n^4}W'_{04},
\end{equation}  
where the expression  $W'_{04}$  is quite cumbersome, and  we do not present it here.

In the results of the  normalization, we received a  new Birkhoffian $B^*$  in the new variables $\tilde{\epsilon}$.
Thus, 
\[
B^*=-\frac{\mu_1^2}{2\tilde{\alpha}^2}-m \frac{\mu_1^2}{\tilde{\alpha}^2}\tilde{\eta}+m^2\frac{\mu_1^2}{4 \tilde{\alpha}^2}\left(1-\frac{3 }{2}\tilde{e}^2+\frac{15}{2}\tilde{e}^2 \cos2\tilde{g}\right)+m^3*0\; +
\]
\[
m^4 \frac{\mu_1^2}{16 \tilde{\alpha}^2}\left[\frac{49}{4}-\frac{873}{4}\tilde{e}^2+\frac{4347}{32}\tilde{e}^4-
   \left(\frac{333}{4}\tilde{e}^2-\frac{237}{8}\tilde{e}^4\right)\cos 2\tilde{g}+\frac{615}{32}\cos 4\tilde{g}\right].
\]

In the example above, for the plane problem, we see that the functions $B^*$ and $W$ are presented in the form of 
truncated series in  $ m=n'/n $, but the coefficients of these series are in  closed form.

The solution  for the spatial restricted three-body problem was obtained in the Mathematica package  up to the fifth order
in the small parameter $m$. The obtained solution coincides with the result of  \cite{Hori_Gen} up to the fifth order.
The Appendix provides an analytical expression for Birkhoffian $B^*$ for the spatial restricted three-body problem.

The new system of equations averaged over $l$ has the form
\begin{eqnarray} \label{am77}
\frac{d\tilde{\alpha}}{dt}&=&0,\ \ \ \frac{d\tilde{l}}{dt}=\frac{1}{\mu_1 }\frac{\partial{B^*}}{\partial{\tilde{\alpha}}}-\frac{\tilde{\eta}}{\mu_1 \tilde{\alpha}}\frac{\partial{B^*}}{\partial{\tilde{\eta}}} ,\nonumber\\
\frac{d\tilde{\eta}}{dt}&=&-\frac{1}{\mu_1\tilde{\alpha}}\frac{\partial{B^*}}{\partial{\tilde{g}}},\nonumber\\
\frac{d\tilde{\gamma}}{dt}&=&\frac{\tilde{\gamma}}{\mu_1\tilde{\alpha}\:\tilde{\eta}}\frac{\partial{B^*}}{\partial{\tilde{g}}}-\frac{1}{\mu_1\tilde{\alpha}\:\tilde{\eta}}\frac{\partial{B^*}}{\partial{\tilde{h}}},\nonumber\\
\frac{d\tilde{g}}{dt}&=&\frac{1}{\mu_1\tilde{\alpha}}\frac{\partial{B^*}}{\partial{\tilde{\eta}}}-\frac{\tilde{\gamma}}{\mu_1\tilde{\alpha} \:\tilde{\eta}}\frac{\partial{B^*}}{\partial{\tilde{\gamma}}},\nonumber\\
\frac{d\tilde{h}}{dt}&=&\frac{1}{\mu_1\tilde{\alpha}  \tilde{\eta}}\frac{\partial{B^*}}{\partial{\tilde{\gamma}}}.
\end{eqnarray}
Equations (\ref{am77}) were obtained from (\ref{am48})  using the following substitution:
\[
\frac{d\:u}{d\:t}=\frac{\alpha^2}{r}\frac{d\:l}{d\:t}-\frac{\alpha^2}{r}\frac{\eta \sin u}{\sqrt{1-\eta^2}}\frac{d\:\eta}{d\:t}.
\]
The resulting system of equations has two degrees of freedom, since the first two equations are separated from the system.

Remark:The function $B^*$ in the Apprndix includes variables without tilde, because it is the result of calculations on 
a computer.
 
\subsection{Representation of analytical solution of restricted three-body problem using the Delaunay arguments.The second example}
\label{sec3.2}

The motion of the satellite due to the attraction of the central planet and the disturbing body $S$ is considered by assuming
that all the bodies are mass-points, and the satellite has infinitesimal mass, the central body has mass $m_{0}$ and the body $S$ has mass $M$.

 For this task, we use the   variables $\epsilon= (\alpha,E,J,\Lambda,D,l.F,l' )$. Variable $\alpha$ was defined
in the previous example;  $\Lambda$ is
an auxiliary variable conjugate to the variable $l'$ (mean anomaly of the disturbing body); $D,\:l,\:F,\:l'$ are Delaunay's
basic arguments: $D=\lambda-\lambda'$, $F=\lambda-\Omega$, $l=\lambda-\pi$, where  $\lambda$ is the orbital longitude of
the satellite, measured from the chosen direction of the $x$-axis of a rectangular coordinate system; $\Omega$ and $\pi$
 are  the longitude of the ascending node and pericenter of the satellite orbit, respectively; $\lambda'$ is the mean
longitude of the perturbing body. Variables $ E $ and $ J $ are determined by the  formulas:
\begin{eqnarray} \label{am78}
E^2=2 (1-\sqrt{1-e^2}),\ \  
J^2\:=4\sqrt{1-e^2}\; \gamma^2,\  \gamma=\sin(i/2),
\end{eqnarray}
where $e$   is the eccentricity satellite's orbit, $i$  is the inclination of the satellite orbit   to the plane in
which the body S moves.

Birkhoffian of this system is represented as follows:
\begin{equation} \label{am79}
B=-\frac{\mu_1^2}{2 \alpha^2}+\nu\:\Lambda + R,
\end{equation}
where $\mu_1$ was defined in the previous example,  $\nu$ is a mean motion of the  disturbing body.
The first term ($-\mu_1^2/2 \alpha^2$) in the expression (\ref{am79}) is  due to the attraction of the planet in the absence
of disturbances. The second term ($\nu\:\Lambda$) is introduced to eliminate the explicit dependence function $B(\epsilon)$
from time, and $R=\nu^2 S'$ is a  disturbing
function. $S'$  has the form of a truncated series
\begin{equation} \label{am80}
S'=\sum_{i,j}A_{i}^{j} \nu^{j_1} \alpha'^{j_2} E^{j_3} J^{j_4} e'^{j_5} \cos(\sin)(i_1 D+i_2 l+i_3 F+i_4 l'),
\end{equation}
where $A_{i}^{j}$ are numerical coefficients; $\alpha'= \alpha/\sqrt{a'}$,  $a'$ and $e'$ are the semi-major axis and the
eccentricity of the orbit on which the disturbing body moves.

Thus, unlike the previous example, here the perturbing function is explicitly dependent on the mean satellite anomaly. 
It follows that all the expressions now include an expansion in powers of the eccentricity of the satellite orbit. 
These are the traditional expansions of the perturbation theory of Celestial Mechanics. As  mentioned above,
the generalized  Lie generator  can be represented  as the   generalized Poisson bracket (\ref{am16}), thus preserving the 
invariant properties of Poisson brackets, without destroying the d'Alembert
characteristics in the expansions of the perturbation theory.

Pfaffian of the problem can be written as
\begin{eqnarray} \label{am81}
\Phi= \alpha \mu_1 (1-\frac{1}{2}E^2-\frac{1}{2}J^2)\:d\: D +\frac{1}{2}\alpha\: \mu_1 E^2 d\:l+\frac{1}{2}\alpha\:\mu_1\:J^2\:d\:F+\nonumber\\
(\Lambda + \alpha\:\mu_1(1-\frac{1}{2}E^2-\frac{1}{2}J^2))\:d\:l'- B\:d\:t.
\end{eqnarray}
Birkhoff's equations are represented in this case as follows:
\begin{eqnarray} \label{am82}
\frac{d\alpha}{dt}&=&-\frac{1}{\mu_1}\left(\frac{\partial{B}}{\partial{D}}+\frac{\partial{B}}{\partial{l}}+
\frac{\partial{B}}{\partial{F}}\right),\nonumber\\
\frac{dE}{dt}&=&\frac{1}{\mu_1\alpha}\left(\frac{E}{2}\frac{\partial{B}}{\partial{D}}+\left(\frac{E}{2}-
\frac{1}{E}\right)\frac{\partial{B}}{\partial{l}}+
\frac{E}{2}\frac{\partial{B}}{\partial{F}}\right),\nonumber\\
\frac{dJ}{dt}&=&\frac{1}{\mu_1\alpha}\left(\frac{J}{2}\frac{\partial{B}}{\partial{D}}+
\frac{J}{2}\frac{\partial{B}}{\partial{l}}+
\left(\frac{J}{2}-\frac{1}{J}\right)\frac{\partial{B}}{\partial{F}}\right),\nonumber\\
\frac{d\Lambda}{dt}&=&\frac{\partial{B}}{\partial{D}}-\frac{\partial{B}}{\partial{l'}},\nonumber\\
\frac{dD}{dt}&=&\frac{1}{\mu_1}\left(\frac{\partial{B}}{\partial{\alpha}}-
\frac{E}{2\alpha}\frac{\partial{B}}{\partial{E}}-\frac{J}{2\alpha}\frac{\partial{B}}{\partial{J}}\right)-
\frac{\partial{B}}{\partial{\Lambda}},\nonumber\\
\frac{dl}{dt}&=&\frac{1}{\mu_1}\left(\frac{\partial{B}}{\partial{\alpha}}+
\left(\frac{1}{\alpha E}-\frac{E}{2\alpha}\right)\frac{\partial{B}}{\partial{E}}-
\frac{J}{2\alpha}\frac{\partial{B}}{\partial{J}}\right),\nonumber\\
\frac{dF}{dt}&=&\frac{1}{\mu_1}\left(\frac{\partial{B}}{\partial{\alpha}}-
\frac{E}{2\alpha}\frac{\partial{B}}{\partial{E}}+\left(\frac{1}{\alpha\:J}-
\frac{J}{2\alpha}\right)\frac{\partial{B}}{\partial{J}}\right),\nonumber\\
\frac{dl'}{dt}&=&\frac{\partial{B}}{\partial{\Lambda}}.
\end{eqnarray}

Birkhoffian $B'_{00}$ unperturbed motion we define as
\begin{equation} \label{am83}
B'_{00}=-\frac{\mu_1^2}{2 \alpha^2}+\nu\:\Lambda+\nu^2R'_{00}.
\end{equation}
The expression  $R'_{00}$ includes only those terms of the secular part of the disturbing function that contain
variables $E$ and $\Theta$ of  no higher than second  degree \cite{Brouwer}:
\begin{equation} \label{am84}
R'_{00}=-\frac{1}{4} \alpha^4-\frac{3}{8}\alpha^4 E^2+\frac{3}{8}\alpha^4 J^2 .
\end{equation}

With this choice of the unperturbed motion of the satellite, the averaging procedure involves all angular 
variables $(D, l, F, l')$, i.e., the system is  a non-degenerate.  The frequencies associated with these variables  
 are determined from the system of equations (\ref{am82}) using the substitution $B=B'_{00}$.

The  analytical expressions for the frequencies of this system are determined by formulas:
\begin{eqnarray} \label{am85}
\omega_1& =& \frac{\mu_1}{\alpha^3}-\frac{\nu^2 \alpha^3}{\mu_1}\left(1+\frac{9}{8} E^2-\frac{9}{8} J^2\right)-\nu,\nonumber\\
\omega_2& =& \frac{\mu_1}{\alpha^3}-\frac{\nu^2 \alpha^3}{\mu_1}\left(1+\frac{9}{8} E^2-\frac{9}{8} J^2\right)-\frac{3}{4}\frac{\alpha^3}{\mu_1},\nonumber\\
\omega_3& =& \frac{\mu_1}{\alpha^3}-\frac{\nu^2 \alpha^3}{\mu_1}\left(1+\frac{9}{8} E^2-\frac{9}{8} J^2\right)+\frac{3}{4}\frac{\alpha^3}{\mu_1},\nonumber\\
\omega_4& =& \nu.
\end{eqnarray}

In this example, the homological equation  is written in a general form as:
\begin{equation} \label{am86}
\sum_{j=1}^4 \omega_j\frac{\partial W_{0k}}{\partial \epsilon_{j+4}}=B_{0k}^*-B_{0k},
\end{equation}
where the functions $W_{0k},\;B_{0k}^*$ and $B_{0k}$  have the same meaning as in the previous example. 

The solution of Eq. (\ref{am86}) leads to the appearance of divisors of the following form:
\begin{eqnarray} \label{am87}
d&=&((i_1+i_2+i_3)\mu_1\alpha ^{-3}+(i_1+i_2+i_3)(-\nu^2 \mu_1^{-1}\alpha^3)+\nonumber\\
 & &(i_1+i_2+i_3)(-9/8\: \nu^2 \mu_1^{-1}\alpha^3\:E^2)+(i_1+i_2+i_3)(9/8\: \nu^2 \mu_1^{-1}\alpha^3 J^2)+\nonumber\\
 & &(i_1-i_4) \nu +(i_2-i_3)(3/4\: \nu^2\:\mu_1^{-1} \alpha^3)).
\end{eqnarray}

From the expression (\ref{am87}) we can obtain conditions for periods of various disturbances.
Let us consider the trigonometric arguments  
\[
i_1 D+i_2 l+i_3 F+i_4 l'
\]
in the expression (\ref{am80}).

Mean anomaly  $l$ of the satellite  is contained  in $ D,\:L,\: F$;  therefore
the condition $i_1+i_2+i_3 \neq 0 $ determines those terms of the perturbing function whose periods  are commensurate with
the period of orbital motion of a satellite around the planet.  This period is denouted via $P^{(l)}$.

For the terms of the perturbing function whose periods  are commensurate with the
orbital period of a planet around the Sun ($P^{(l')}$), the conditions are  $i_1+i_2+i_3 = 0, i_1 \neq i_4 $.
These terms include the mean anomaly $l'$ of  the planet,  but do not contain the mean anomaly $l$ of a satellite.

The conditions  for the long-period terms would be as follows: $i_1+i_2+i_3 = 0,\  i_1 = i_4,\  i_2\neq i_3 $. These terms 
contain neither $l$ nor $l'$.

The  system under consideration   is non-resonant; therefore, in our case,
the perturbation  function $R$ may be decomposed into its  $P^{(l)}$-period, $P^{(l')}$-period  and long-period  parts.
Next, we  can carry out the operation of averaging over different periods. 
Normalization of this kind has already been used by several authors\cite{Hori_Gen, Deprit_Henrard}.

In  averaging the perturbing function over mean anomaly $l$, we can take  a Keplerian motion
as the unperturbed motion of the satellite. Thus, we have from (\ref{am83}) that
\[
B'_{00}=B_{00}^{(l)}=-\frac{\mu_1^2}{2\alpha^2}
\]
and $\omega_1=\omega_2=\omega_3=\mu_1 \alpha^{-3}$.

Therefore, the homological equation at this stage can be written as
\begin{eqnarray} \label{am88}
-\frac{\mu_1}{\alpha^3}\left(\frac{\partial{W_{0k}}}{\partial{D}}+\frac{\partial{W_{0k}}}{\partial{l}}+
\frac{\partial{W_{0k}}}{\partial{F}})\right)=B_{0k}^*-B_{0k}.
\end{eqnarray}
This equation is easily solved. Function $B_{0k}^*-B_{0k}$  contains only those terms for which $i_1+i_2+i_3 \neq 0$.
It follows that we can obtain the function $W_{0k}$  from the expression $B_{0k}^*-B_{0k}$ with the help of the 
substitution $\cos \rightarrow -\sin$, and  then  multiplying  the result by 
 
\[
 \frac{1}{(i_1+i_2+i_3) \mu_1 \alpha^{-3}}.
\]

Next, we form the k-th column of the generating matrix $\Psi$, as  described in the previous example. When the
 $\Psi$ matrix has been defined up to the desired order, we can  begin the process of the transformation of variables. 
 
 For the transformation of variables, we enter on the first line of the triangular matrix (\ref{am58}) the following information:
$B_{00}=\epsilon_i, B_{01}=0,\ldots,B_{0m}=0 $.  We then use the above-referenced algorithm  (\ref{am60}). To obtain the inverse
 transformation, we must reverse the signs of all elements of the $\Psi$ matrix to the opposite signs. The result is the following 
transformation of variables: $\epsilon \leftrightarrow \hat{\epsilon}$.

At  the first step of normalization, we obtain the integral of motion  $\hat{\alpha}=const.$ It follows that in oder to   eliminate  the $P^{(l')}$-terms from perturbing function, we can use  the function $B'_{00}$ in the form
\[
B'_{00}=B_{00}^{(l')}=\nu \Lambda.
\]
 In this case the frequencies are defined so that  $\omega_1=-\nu,\ \omega_4=\nu$,
and the homological equation is
\begin{eqnarray} \label{am89}
\nu\left(\frac{\partial{\hat{W}_{0k}}}{\partial\hat{D}}-\frac{\partial{\hat{W}_{0k}}}{\partial{\hat{l'}}}\right)=\hat{B}_{0k}^*-\hat{B}_{0k}.
\end{eqnarray}

It follows from expression (\ref{am87}) that in order to  eliminate   the long-period perturbations from the  Birkhoffian, one can use the homological equation in the form :
 \begin{eqnarray} \label{am90}
\frac{3}{4}\nu^2\frac{\tilde{\alpha}^3}{\mu_1}\left(\frac{\partial{\tilde{W}_{0k}}}{\partial{\tilde{l}}}-\frac{\partial{\tilde{W}_{0k}}}{\partial{\tilde{F}}}\right)=\tilde{B}_{0k}^*-\tilde{B}_{0k}.
\end{eqnarray}

Using this algorithm, we can obtain any desired order of transformation. As a result,    
Eq. (\ref{am82})
is transformed into the equivalent system:
\begin{eqnarray} \label{am91}
\frac{d\tilde{\alpha}}{dt}&=&0,\nonumber\\ 
\frac{d\tilde{E}}{dt}&=&0, \nonumber\\
\frac{d\tilde{J}}{dt}&=&0, \nonumber\\
\frac{d\tilde{\Lambda}}{dt}&=&0,\nonumber\\
\frac{d\tilde{D}}{dt}&=&\frac{1}{\mu_1}\left(\frac{\partial{\tilde{B}}}{\partial{\tilde{\alpha}}}-
\frac{\tilde{E}}{2\tilde{\alpha}}\frac{\partial\tilde{{B}}}{\partial{\tilde{E}}}-\frac{\tilde{J}}{2\tilde{\alpha}}\frac{\partial{\tilde{B}}}{\partial{\tilde{J}}}\right)-
\frac{\partial{\tilde{B}}}{\partial{\tilde{\Lambda}}}=\tilde{n}-\nu,\nonumber\\
\frac{d\tilde{l}}{dt}&=&\tilde{n}+\frac{1}{\tilde{\alpha }\tilde{E}}\frac{\partial{\tilde{B}}}{\partial{\tilde{E}}},\nonumber\\
\frac{d\tilde{F}}{dt}&=&\tilde{n}+\frac{1}{\tilde{\alpha}\:\tilde{J}}\frac{\partial{\tilde{B}}}{\partial{\tilde{J}}},\nonumber\\
\frac{d\tilde{l'}}{dt}&=&\nu.
\end{eqnarray}
The solutions are trivial:
\begin{eqnarray} \label{am92}
\tilde{\alpha}&=&\alpha_0,\  \tilde{E}=E_0,\  \tilde{J}=J_0,\  \tilde{\Lambda}=\Lambda_0, \nonumber\\
\tilde{D}&=&(\tilde{n}-\nu)t+D_0,\ \tilde{l}=(\tilde{n}-\frac{d\tilde{\pi}}{dt})t+l_0,\nonumber\\
\tilde{F}&=&(\tilde{n}-\frac{d\tilde{\Omega}}{dt})t+F_0,\ l'=\nu t+l'_0.
\end{eqnarray}
Here $\alpha_0,E_0,J_0,D_0,l_0,F_0$ are the constants of the analytical theory, which are the mean
elements of the orbit of the satellite  in the epoch $t_0$, $l'_0$ is the mean anomaly of the disturbing body in reference to the
same moment $t_0$, $\Lambda_0$ is an auxiliary parameter, which is not present in the final expansions,
and $t$ is the time  in Julian days from the epoch $t_0$. 
The mean motion of the longitude $\tilde{n}$, of the longitude of the pericentre $d\tilde{\pi}/dt$ and of the longitude of the node
$d\tilde{\Omega}/dt$ are represented by series of the form
\[
C_q=\sum_j K_j m^{j_1}\alpha_0^{j_2} E_0^{j_3} J_0^{j_4} e^{'j_5}.
\]
The mean elements of the orbit of the satellite are obtained using the expansions for the inverse transformation 
of variables.

 We have used a similar algorithm in an  earlier work \cite{Boronenko_Shmidt}. The literal solution of the restricted
three-body problem, which the authors obtained  up to the 11-th order with respect to the minor parameter $m=\nu/n$,
was applied to the investigation  of motion of Phoebe, the ninth satellite of Saturn. In this article, we  derived
the algorithm in the  context of the theory of Birkhoff.  A more complete description of the solution of this problem can be 
found in \cite {Boronenko_Shmidt}. 

\section{Conclusion}
\label{sec4} 

In this article, we demonstrated the usefulness of the Lie transformation  algorithm for Birkhoff systems, which are  described 
by the equations of the following form:
\[
\dot{\theta}_p =\sum_{s=1}^n a_{ps}(\theta)\frac{\partial B(\theta)}{\partial\theta_s},\ \ (p=1,\ldots.n),
\]
 where  the function $B(\theta)$ is the Hamiltonian  expressed in the special variables $\theta$ of the phase space.
 Indeed, Birkhoff's  autonomous  equations  are  the Hamilton equations, which are  represented  in non-canonical variables
 in the phase space.  However, we use 
the term 'Birkhoffian' because of certain physical differences  with Hamiltonian, in that  the matrix $(a_{ps})$
contains the  time t via the variables $\theta(t)$. A more detailed discussion can be found in \cite{Santilli}. 

 Tensor $a_{ps}$  is the Birkhoff tensor,  expressed in terms of 
the Poisson brackets.
 In accordance with the terminology  of \cite{Santilli}, $a_{ps}$  is the contravariant  Birkhoff tensor, also called
 the Lie tensor.

In Sect. 2.2 we showed that the representation of the Lie generator  for canonical systems in terms of 
 the special variables of the phase space led to its expression   through the Birkhoff tensor $a_{ps}$.  
 As shown by \cite{Santilli}, the Birkhoff tensor $a_{ps}$ and its associated symplectic form $\Omega$ preserve their Lie 
and symplectic character under arbitrary transformations. This allowed us to
 use  the autonomous Birkhoff equations in the construction of the Lie series
perturbation theory.

The basis of the algorithm is a  generalized Lie generator, which we express in terms of the tensor $a_{ps}$.
To reduce the need for cumbersome operations of multiplication of series, we introduced
the generating  matrix $\Psi$, which  is defined with the help of the tensor $a_{ps}$  and  partial derivatives of the generating
function. 
The matrix $\Psi$ is also used  for the direct and inverse coordinate transformations.

In this work, we have demonstrated the algorithm,  based on a generalized Lie generator, using two examples from
Celestial Mechanics.

In the first  example, we considered the satellite case of the spatial restricted three-body problem,
using an  averaging method, based on a Lie  transformation of the Birkhoffian system.
The  new Birkhoffian,   averaged over  mean anomaly of a  satellite, was obtained in the form of  series in $ m $
(ratio of mean motions of the Sun  and satellite), but in closed form with respect to eccentricity and inclination.
The accuracy of the  analytical expansion is O $(m^5)$. Our results were coincident  with the result of  \cite{Hori}
up to the fifth order \cite{Boronenko}.

In the second example, we  represented an  analytical solution of restricted three-body problem using 
the Delaunay arguments $(D, l, F, l')$. Unlike the previous example, here we dealt with an explicit expression of
the perturbing function in terms of
the  mean anomaly of the satellite. Therefore, all the considered expressions now   included power series in the eccentricity 
of the satellite orbit. 
 These are the traditional expansions of the perturbation theory of Celestial Mechanics. This example shows that
 the Lie generator, expressed in terms of  the Birkhoff tensor, preserves the invariant properties of
Poisson brackets. For example, the use of the generalized Lie generator  does not  destroy the d'Alembert characteristic 
in the series of   perturbation theory.  We used a similar
algorithm in an earlier work  for constructing an analytical theory of motion of Phoebe, the ninth satellite
 of Saturn \cite{Boronenko_Shmidt}.  In this paper, we derived the algorithm in accordance with the theory Birkhoff: we  introduced 
the Pfaffian, Birkhoff's
equations, and the generalized Lie generator  for this problem. A more complete description  of the problem can be 
found in \cite {Boronenko_Shmidt}. 

The  examples above show that the proposed algorithm does not violate the basic approaches of the standard 
Lie transformation theory, but it provides an  efficient alternative in the case where there is a need to use  the non-canonical 
 variables $(\theta)$ in phase space.  
 Here, the Birkhoffian scheme provides a clear way in which to build
the solution. It is important that all operations are performed only in the variables $(\theta)$. As shown by the examples,
 the proposed algorithm does not increase the number of operations compared to the standard Lie transformation theory,
 but in the case of non-canonical variables, the generating matrix $\Psi$ allows  the number of multiplications
 of large expressions to be reduce.

In addition to the technical characteristics of the algorithm, the properties of Birkhoff systems
 are useful for developing  a common approach  to the process of formulating  a problem. 
 The use of 1-forms allows us to expand the types of coordinate transformations, as the dynamic Pfaffians  
 can be represented by a large number  of different forms.   Pfaffian does not change its form when a transformation of
coordinates  is made in the phase space with dimension $2n+1$, i.e., in the extended phase space of Pfaff.  
Therefore, in our work, we define   Birkhoff functions as a components  of the Pfaff vector. 
 
  It should be noted that the Birkhoff tensor $\Omega_{ps}$ in covariant form coincides with
the matrix of the  Lagrange brackets.  Using Keplerian elements  as coordinates in  phase space is a simple way  of deriving  the Lagrange planetary equations. In this case, the Birkhoff autonomous equations 
coincide with Lagrange's planetary equations, and the method can be used 
for analytical integration of these  equations.

 Unlike the other Lie  transformations algorithms for non-canonical systems (see e.g., \cite{Nayfeh}), the Lie series transformations for  the autonomous  Birkhoff equations that we have considered here are  derived from the Pfaff-Birkhoff
 variational 	principle, which is more general than the Hamilton principle.
 The use of 1-form in the formulation of equations of motion in dynamics
 renders the Birkhoff  method more universal and flexible.The Birkhoff equations  have a tensorial  character; therefore, their 
form is independent of the coordinate system that is used. 
 
\section*{Acknowledgement}

The author expresses sincere gratitude to the head of the Department of theoretical physics, Tomsk state pedagogical University,
Professor I. L. Buchbinder for consultations and discussions, and Professor of theoretical physics V. Ya. Epp for his
suggestions that helped to improve the presentation of the article.

\appendix
\section{The expression for the averaged Birkhoffian (the first example)}

The resulting averaged    Birkhoffian  is represented as follows:

\[
B^*=B_{00}^*+B_{01}^*+B_{02}^*+B_{03}^*+B_{04}^*+B_{05}^*,\ \ B_{00}^*=B_{00},\ \ B_{01}^*= B_{01},
\]
where
\begin{eqnarray*}
B_{02}^*&=&\frac{1}{16}\nu^2 a^2 (((2+3 e^2)(1-3 \gamma^2+3(-1+\gamma^2)\cos(2 h))-\\
        & &15 e^2 \cos(2 g)(1-\gamma^2+(1+\gamma^2) \cos(2 h)+30 e^2\gamma \sin(2 g) \sin(2 h)),
\end{eqnarray*}
\begin{eqnarray*}				
B_{03}^*&=&0, 
\end{eqnarray*}
\begin{eqnarray*}
B_{04}^*&=&\frac{1}{4096}\frac{\nu^4 a^2}{n^2}(8(47+282 \gamma^2 +63\gamma^4)+63 e^4 (239+170\gamma^2+143 \gamma^4)-\\
        & & 72 e^2 (377+190\gamma^2+209 \gamma^4)+2592 \cos(2 h) +168 \cos(4 h)-24 e^2 \cos(2 g) (1-\gamma^2+\\
        & &(1+\gamma^2) \cos(2 h)(27(2+e^2)+5(78-37 e^2)\gamma^2+5(-78+37 e^2)(-1 +\gamma^2) \cos(2 h)-\\
        & &410 e^2\gamma\sin(2 g)\sin(2 h)+3((56\gamma^2(-2 +\gamma^2)-1672 e^2 (-1+\gamma^2)^2 +\\
        & &1001 e^4(-1 +\gamma^2))\cos(4 h)+205 e^4 \cos(4 g) (-3(-1 +\gamma^2)^2+4(-1+\gamma^4)\cos(2 h)-\\
        & &(1+6\gamma^2+\gamma^4)\cos(4 h))+16 e^2\gamma(27(2+e^2)+\\
        & & 5(78-37 e^2)\gamma^2)\sin(2 g)\sin(2 h)+4\cos(2 h)(-8\gamma^2(20+7\gamma^2)+\\
        & &152 e^2(-13+2\gamma^2+11\gamma^4)-7 e^4(-199+56\gamma^2+143\gamma^4)+\\
        & & 20 e^2(-78+37 e^2)\gamma(-1+\gamma^2)\sin(2g)\sin(2h))),
\end{eqnarray*}
\begin{eqnarray*}
B_{05}^*&=&\frac{1}{128}\frac{\nu^4 a^2}{n^2}m\:\eta (\gamma(176-2775 e^2+870 e^4 +(212-1895 e^2+675 e^4)\gamma^2-\\
        & &(212-1895 e^2+675 e^4)(-1+\gamma^2)\cos(2 h)+4 e^2(-101 +17e^2)\gamma \cos(2 g)(3-\\
        & &3\gamma^2+(-1+3\gamma^2)\cos(2 h))-8e^2(-101+17 e^2)(-1 +2 \gamma^2)\sin(2 g) \sin(2 h)).
\end{eqnarray*}
  
The above expressions were checked by comparison with the results
 of (\cite{Hori_Gen}), under the condition $\gamma = 1$ and $h = 0$.
 Complete coincidence was found with the analytical expressions for the functions $B_{02}^*$, $B_{03}^*$, $B_{04}^*$.
 For the function $B_{05}^*$,  only the secular part coincided. This discrepancy can be explained by the use of different methods ( Lie transformations and von Zeipel method) in solvving the problem. 

Note. Expressions for the $B_{0i}$ functions in the traditional form  can be obtained 
in the   Mathematica package using the function  $Expand[TrigReduce[B_{0i}]]$. Examples of these expressions
for the planar version of the problem can be found in section 3.1


\end{document}